# Cross-Correlation Earthquake Precursors in the Hydrogeochemical and Geoacoustic Signals for the Kamchatka Peninsula

G. V. Ryabinin[1], V. A. Gavrilov[2], Yu. S. Polyakov[3], and S. F. Timashev[4,5]

[1]Kamchatka Branch, Geophysical Survey of Russian Academy of Sciences, Petropavlovsk-Kamchatsky, Russia; e-mail: gena@emsd.iks.ru

[2]Institute of Volcanology and Seismology, Far Eastern Branch of the Russian Academy of Sciences, Petropavlovsk-Kamchatsky, Russia; e-mail: vgavr@kscnet.ru

[3]USPolyResearch, Ashland, PA, USA; e-mail: ypolyakov@uspolyresearch.com (corresponding author)

[4]Institute of Laser and Information Technologies, Russian Academy of Sciences, Troitsk, Moscow Region, Russia; e-mail: serget@mail.ru

[5]Karpov Institute of Physical Chemistry, Moscow, Russia

A b s t r a c t

We propose a new type of earthquake precursor based on the analysis of correlation dynamics between geophysical signals of different nature. The precursor is found using a two-parameter cross-correlation function introduced within the framework of flicker-noise spectroscopy, a general statistical physics approach to the analysis of time series. We consider an example of cross-correlation analysis for water salinity time series, an integral characteristic of the chemical composition of groundwater, and geoacoustic emissions recorded at the G-1 borehole on the Kamchatka peninsula in the time frame from 2001 to 2003, which is characterized by a sequence of three groups of significant seismic events. We found that cross-correlation precursors took place 27, 31, and 35 days ahead of the strongest earthquakes for each group of seismic events, respectively. At the same time, precursory anomalies in the signals themselves were observed only in the geoacoustic emissions for one group of earthquakes.

______________________________________________________________

© AG template

## 1. INTRODUCTION

Earthquake prediction within a time frame of several months to less than an hour before the catastrophic event, which is often referred in literature as "short-term" prediction, has been a subject of extensive research studies and controversial debates, both in academia and mass media, in the past two decades (Geller 1997, Geller *et al.* 1997, Wyss *et al.* 1997, Uyeda *et al.* 2009, Cicerone *et al.* 2009). One of the key areas in this field is the study of earthquake precursors, physical phenomena that reportedly precede at least some earthquakes. The precursory signals are usually grouped into electromagnetic, hydrological/hydrochemical, gasgeochemical, geodetic, and seismic (Geller 1997, Hartmann and Levy 2005, Uyeda *et al.* 2009, Cicerone *et al.* 2009, Ryabinin *et al.* 2011).

Despite the large number of earthquake precursors reported in literature, most of which are summarized by Hartmann and Levy (2005), Cicerone *et al.* (2009), an International Commission on Earthquake Forecasting for Civil Protection concluded on 2 October 2009, "the search for precursors that are diagnostic of an impending earthquake has not yet produced a successful short-term prediction scheme" (ICEFCP 2009). The reports of the International Association of Seismology and Physics of the Earth's Interior contain similar findings (Wyss and Booth 1997). The lack of confidence can be attributed to several reasons. First, some fundamental aspects of many non-seismic signals, for example, lithosphere-atmosphere-ionosphere coupling and propagation of high-frequency electromagnetic signals in the conductive earth, are unresolved, and many of the proposed physical models are questionable (Uyeda *et al.* 2009). Second, the experimental data on precursory signals are often limited to few earthquakes and few measurement sites and they frequently contain gaps and different types of noise (Hartmann and Levy 2005, Cicerone *et al.* 2009, Uyeda *et al.* 2009). Third, different techniques of identifying the anomalies are used for different signals or even in different studies for the same signal. In some cases, the anomalous changes are determined by analyzing the signals themselves (Hartmann and Levy 2005, Uyeda *et al.* 2009, Cicerone *et al.* 2009), while in other cases they are identified by studying the derived statistics or functions, such as Fisher information or scaling parameters (Telesca *et al.* 2009a,b). Moreover, seasonal changes and instrumentation or other background noise often need to be filtered out prior to the identification of precursors.

In view of the above, we believe that earthquake precursor research can be advanced by employing a phenomenological approach to the analysis of signals of different types in the same local geographic region. We assume that a large earthquake may be preceded by a reconfiguration of a geophysical system on different time and space scales, which manifests itself in qualitative changes of various signals within relatively short time intervals. In our previous study (Ryabinin *et al.* 2011), we used a nonstationarity factor introduced within the framework of flicker-noise spectroscopy (FNS), a statistical physics approach to the analysis of time series (Timashev and Polyakov 2007, Timashev 2007, Timashev *et al.* 2010b), to determine the time moments of abrupt rearrangements in the seismic zone of the Kamchatka peninsula. We analyzed together the nonstationarity factors for chlorine-ion concentration of groundwater and geoacoustic emissions in a deep borehole within that seismic zone in the time frame around a large earthquake on 8 October 2001. Our analysis showed that nonstationarity-factor peak values (potential precursors) take place in the interval from 70 to 50 days before the earthquake for the hydrogeochemical data and at 29 and 6 days in advance for the geoacoustic data.

In this paper, we suggest a new type of earthquake precursor based on cross-correlation analysis of the dynamics for two different types of signals: hydrogeochemical and geoacoustic. The idea of examining qualitative changes in the cross-correlations for signals of different types stands on the systematic analysis of common earthquake precursors presented by Cicerone *et al.* (2009). Most of the known models suggest that earthquake precursory anomalies are driven by rapid deformations and strain changes within the earth in the rock near or in the fault zone at the region of eventual earthquake rupture. It was shown that the rapid deformations just prior to fracture combined with changes in the groundwater and gas flow in the earth caused by the variation of porosity and permeability in the rock volume, which in its turn is effected by micro-fractures, can generate all of the earthquake precursors studied by Cicerone *et al.* (2009). In other words, cross-correlation precursors rely on a common mechanism of "preparatory" changes attributed to a system reconfiguration preceding a large earthquake. It should be pointed out that the analysis of cross-correlations may uncover hidden qualitative changes which cannot be determined from the analysis of specific signals by themselves because of seasonal changes and instrumentation and other background noise.

We would like to note that the precursory role of synchronization effects in geophysical signals of different nature was previously studied by Lyubushin *et al.* (1997) and Lyubushin (1998, 2000, 2007). Based on the assumption that a large earthquake is preceded by a substantial increase in the synchronization between various geophysical processes, the author devel-oped mathematical methods to build an aggregate signal that provides maximum information on the most general variations common to all analyzed processes. This assumption came from the catastrophe theory stating that the spatial radius of fluctuations increases and the collective component in the behavior of different parts of a complex system rises when the underlying complex system approaches a catastrophe or a phase transition. The aggregate signal used for detecting precursory anomalies is obtained using covariance matrices for multidimensional vectors and Fourier or wavelet transformations (Lyubushin 2007). In contrast to this method, the cross-correlation precursor suggested in this paper is focused on local dynamic correlations for a pair of signals, with zero or non-zero time shift between each other. The method proposed in this paper does not automatically assume that maximum cross-correlations correspond to a precursor, but looks for significant changes in the cross-correlation function, which are believed to be associated with system reconfiguration.

The paper is structured as follows. In Section 2, we provide the fundamentals of FNS and present the cross-correlation function. Section 3 describes the experimental setup. Section 4 discusses the cross-correlation analysis of hydrogeochemical and geoacoustic data for the Kamchatka peninsula in the time frame from 2001 to 2003. Section 5 presents the conclusions.

## 2. FNS CROSS-CORRELATION FUNCTION

Here, we will only deal with the basic FNS relations needed to understand the cross-correlation function. FNS is described in more detail elsewhere (Timashev 2006, Timashev and Polyakov 2007, Timashev 2007, Timashev and Polyakov 2008, Timashev *et al.* 2010)

### 2.1 Principles of flicker-noise spectroscopy

In FNS, all introduced parameters for signal $V(t)$, where $t$ is time, are related to the autocorrelation function

$$\psi(\tau) = \left\langle V(t)V(t+\tau) \right\rangle_{T-\tau} \tag{1}$$

where $\tau$ is the time lag parameter ( $0 \leq \tau \leq T_M$ ) and $T_M$ is the upper bound for $\tau$ ( $T_M \leq T/2$ ). This function characterizes the correlation in the values of dynamic variable $V$ at higher, $t + \tau$, and lower, $t$, values of the argument. The angular brackets in relation (1) stand for the averaging over time interval $[0, T - \tau]$

$$\langle(...)\rangle_{T-\tau} = \frac{1}{T-\tau}\int_0^{T-\tau}(...)dt. \qquad (2)$$

As the length of the averaging interval should be less than the overall length of the time series, $T$, and be equal to $T - |\tau|$, we consider, for simplicity, only nonnegative values of $\tau$. The averaging over interval $[0, T-\tau]$ implies that all the characteristics that can be extracted by analyzing functions $\psi(\tau)$ should be regarded as average values on this interval. To extract the information contained in $\psi(\tau)$ ($\langle V(t)\rangle = 0$ is assumed), the following transforms, or "projections", of this function are analyzed: cosine transforms ("power spectrum" estimates) $S(f)$, where $f$ is the frequency,

$$S(f) = 2\int_0^{T_M} \langle V(t)V(t+t_1)\rangle_{T-\tau} \cos(2\pi f t_1) dt_1 \qquad (3)$$

and its difference moments (Kolmogorov transient structure functions) of the second order $\Phi^{(2)}(\tau)$

$$\Phi^{(2)}(\tau) = \langle [V(t) - V(t+\tau)]^2 \rangle_{T-\tau}. \qquad (4)$$

Here, we use the quotes for power spectrum because according to the Wiener-Khinchin theorem the cosine (Fourier) transform of autocorrelation function is equal to the power spectral density only for wide-sense stationary signals at infinite integration limits.

The information contents of $S(f)$ and $\Phi^{(2)}(\tau)$ are generally different, and the parameters for both functions are needed to solve parameterization problems. By considering the intermittent character of signals under study, interpolation expressions for the stochastic components $S_s(f)$ and $\Phi_s^{(2)}(\tau)$ of $S(f)$ and $\Phi^{(2)}(\tau)$, respectively, were derived using the theory of generalized functions by Timashev (2006). It was shown that structural functions $\Phi_s^{(2)}(\tau)$ are formed only by jump-like (random-walk) irregularities corresponding to a dissipative process of anomalous diffusion, and functions $S_s(f)$, which characterize the "energy side" of the process, are formed by spike-like (inertial) and jump-like irregularities. It should be noted that $\tau$ in Eqs. (1)-(4) is considered as a macroscopic parameter exceeding the sampling period by at least one order of magnitude. This constraint is required to derive the expressions and separate out contributions of dissipative jump-like and inertial (non-dissipative) spike-like components.

## 2.2 Cross-correlation function

FNS cross-correlation expressions allow one to analyze cause-and-effect relations in different signals measured simultaneously. The information about the dynamics of correlation links in variables $V_i(t)$ and $V_j(t)$, where indices $i$ and $j$ denote two different signals, can be extracted by analyzing the dynamics of various correlators. Here, we will limit our attention to the simplest two-point cross-correlation expression characterizing the links between $V_i(t)$ and $V_j(t)$ (Timashev 2006, Timashev and Polyakov 2007, Timashev 2007):

$$q_{ij}(\tau, \theta_{ij}) = \left\langle \left[ \frac{V_i(t) - V_i(t+\tau)}{\sqrt{\Phi_i^{(2)}(\tau)}} \right] \left[ \frac{V_j(t+\theta_{ij}) - V_j(t+\theta_{ij}+\tau)}{\sqrt{\Phi_j^{(2)}(\tau)}} \right] \right\rangle_{T-\tau-|\theta_{ij}|} \quad (5)$$

where $\tau$ is the "lag" time corresponding to different time scales; $\theta_{ij}$ is the "time shift" parameter. Higher values of $\tau$ correspond to coarser (low-resolution) analysis of cross-correlations.

In discrete form, eq. (5) is written as

$$q_{ijd}(n_\tau, n_\theta) = \frac{\sum_{k=U[-n_\theta]|n_\theta|+1}^{N-n_\tau-U[n_\theta]|n_\theta|} \left[ V_{id}(k) - V_{id}(k+n_\tau) \right] \left[ V_{jd}(k+n_\theta) - V_{jd}(k+n_\theta+n_\tau) \right]}{\sqrt{\sum_{k=U[-n_\theta]|n_\theta|+1}^{N-n_\tau-U[n_\theta]|n_\theta|} \left[ V_{id}(k) - V_{id}(k+n_\tau) \right]^2}} \times \frac{1}{\sqrt{\sum_{k=U[-n_\theta]|n_\theta|+1+n_\theta}^{N-n_\tau-U[n_\theta]|n_\theta|+n_\theta} \left[ V_{jd}(k) - V_{jd}(k+n_\tau) \right]^2}}, \quad (6)$$

where

$$n = \lfloor \tau/\Delta t \rfloor, \quad n_\theta = \lfloor \theta_{ij}/\Delta t \rfloor, \quad U[x] = \begin{cases} 1, & x \geq 0; \\ 0, & x < 0. \end{cases}$$

Here, subscript $d$ is used to denote the discrete form and $\Delta t$ is the sampling period.

The dependence of cross-correlation $q_{ij}(\tau, \theta_{ij})$ on $\theta_{ij}$ generally describes the cause-and-effect relation, i.e., the characteristic time of information transfer between points (or events) $i$ and $j$. For instance, it may describe the "flow direction" between signals $V_i(t)$ and $V_j(t)$. When $\theta_{ij} > 0$, event $j$ follows event $i$ or, alternatively, the flow moves from point $i$ to point $j$; when $\theta_{ij} < 0$, it is the opposite. When the distance between points $i$ and $j$ is fixed, the value of $\theta_{ij}$ can be used to estimate the rate of information transfer between these two points. The dependence of the value and magnitude of cross-correlation $q_{ij}(\tau, \theta_{ij})$ on $\tau$ and $\theta_{ij}$ can be used to analyze the flow dynamics with signals $V_i(t)$ and $V_j(t)$ changing in phase ($q_{ij} > 0$) and in antiphase ($\theta_{ij} < 0$). In some cases, both signals may be affected by a common external factor manifesting itself after different time lags.

The magnitude and behavior of the two-parameter expression (5) may significantly depend on the value of selected averaging interval $T$ and upper-bound values of $\tau$ and $\theta_{ij}$, which we will refer to as $\tau_{max}$ and $\theta_{max}$. From the statistical reliability point of view, we set the constraint $\tau_{max} + |\theta_{max}| \leq T/2$.

For conciseness, from now on we will refer to $q_{ij}$ as $q$ and $\theta_{ij}$ as $\theta$.

## 3. DATA

The data were recorded in the south-eastern part of the Kamchatka peninsula located at the Russian Far East. The eastern part of the peninsula is one of the most seismically active regions in the world. The area of highest seismicity localized in the depth range between 0 and 40 km represents a narrow stripe with a length of approximately 200 km along the east coast of Kamchatka, which is bounded by a deep-sea trench on the east (Fedotov 1985).

Specialized measurements of groundwater characteristics were started in 1977 to find and study possible hydrogeochemical precursors of Kamchatka earthquakes. Currently, the observation network includes four stations in the vicinity of Petropavlovsk-Kamchatsky (Fig. 1). The Pinachevo station includes five water reservoirs: four warm springs and one borehole GK-1 with the depth of 1,261 m. The Moroznaya station has a single borehole No. 1 with the depth of 600 m. The Khlebozavod station also includes a single borehole G-1 with the depth of 2,540 m, which is located in Petropavlovsk-Kamchatsky. The Verkhnyaya Paratunka station comprises four boreholes (GK-5, GK-44, GK-15, and GK-17) with depths in the range from 650 to 1208 m.

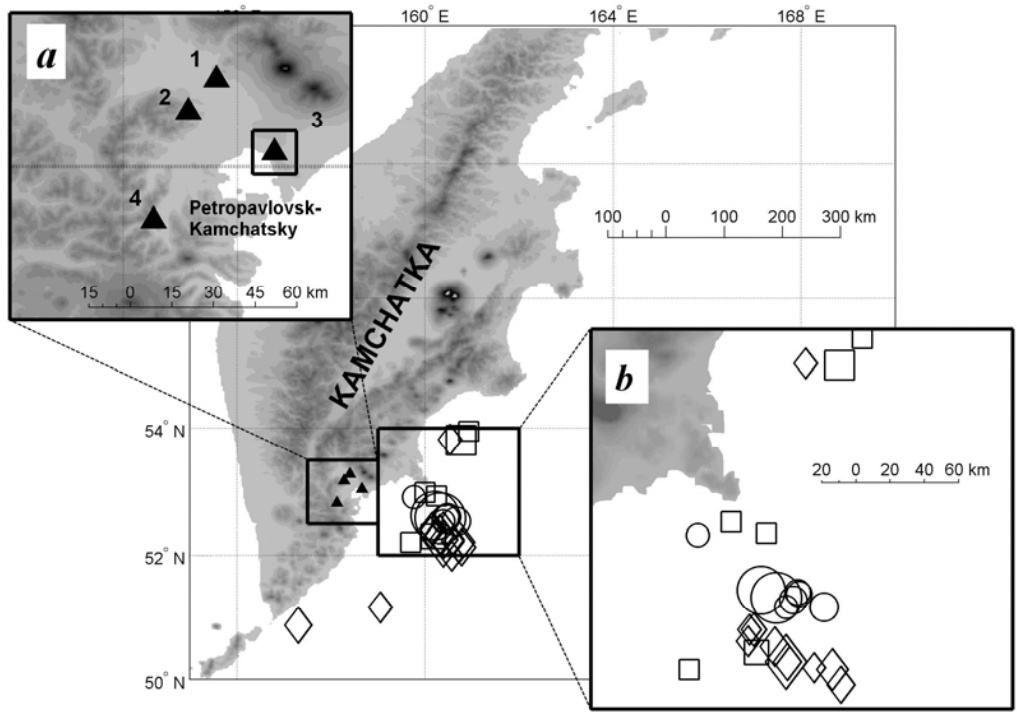

Fig. 1. Schematic of the measurement area, observation points, and epicenters of largest earthquakes ($M \geq 5.0$, $H \leq 50$ km, $D \leq 350$ km) from 2001 to 2003, where $M$ – earthquake magnitude, $H$ – depth, $D$ – distance from the epicenter to the Khlebozavod station. Frame *a* shows an enlarged view of the positions of hydrogeological stations: 1 – Pinachevo, 2 – Moroznaya, 3 – Khlebozavod, 4 – Verknyaya Paratunka. Frame *b* shows an enlarged view of the area including most of the earthquakes. Circles, squares, and diamonds denote earthquakes in 2001; 2002, and 2003, respectively. The earthquakes were selected using the catalog of U.S. Geological Survey: http://earthquake.usgs.gov/earthquakes/eqarchives/epic/.

The system of hydrogeochemical observations includes the measurement of atmospheric pressure and air temperature, measurement of water discharge and temperature of boreholes and springs, collection of water and gas samples for their further laboratory analyses. For water samples, the following parameters are determined: pH; ion concentrations of chlorine ($Cl^-$), bicarbonate ($HCO_3^-$), sulfate ($SO_4^{2-}$), sodium ($Na^+$), potassium ($K^+$), calcium ($Ca^{2+}$), and magnesium ($Mg^{2+}$); concentrations of boric ($H_3BO_3$) and silicone ($H_4SiO_4$) acids. For the samples of gases dissolved in water, the following concentrations are determined: methane ($CH_4$), nitrogen ($N_2$), oxygen ($O_2$), carbon dioxide ($CO_2$), helium (He), hydrogen ($H_2$), hydrocarbon gases: ethane ($C_2H_6$), ethylene ($C_2H_4$), propane ($C_3H_8$), propylene ($C_3H_6$), butane ($C_4H_{10}n$), and isobutane ($C_4H_{10}i$). The data are recorded at nonuniform sam-pling intervals with one dominant sampling frequency. For the Pinachevo, Moroznaya, and Khlebozavod stations, this average sampling frequency is one measurement per 3 days; for the Verkhnyaya Paratunka station, one measurement per 6 days. Multiple studies of the hydrogeochemical data and corresponding seismic activity for the Kamchatka peninsula reported anomalous changes in the chemical and/or gas composition of groundwater prior to several large earthquakes (Kopylova *et al.* 1994, Bella *et al.* 1998, Biagi *et al.* 2000, Biagi *et al.* 2006; Khatkevich and Ryabinin 2006).

Table 1. Largest earthquakes ($M \geq 5.0$, $H \leq 50$ km, $D \leq 350$ km) from 30.06.2001 to 1.06.2003 according to the catalog of U.S. Geological Survey (http://earthquake.usgs.gov/earthquakes/eqarchives/epic/), where $H$ – depth, $D$ – distance from the epicenter to the Khlebozavod station. The earthquake magnitude $M$ is listed using different scales: Mw – moment magnitude, Ms – surface wave magnitude, mb – body wave magnitude.

| Date (dd/mm/yyyy) | LAT | LONG | D (km) | M | | H (km) |
|---|---|---|---|---|---|---|
| 07/10/2001 | 52.62 | 160.47 | 33 | 5.2 | MwHRV | 134 |
| 08/10/2001 | 52.58 | 160.44 | 33 | 5.3 | MwHRV | 133 |
| 08/10/2001 | 52.59 | 160.32 | 48 | 6.5 | MwHRV | 126 |
| 08/10/2001 | 52.63 | 160.21 | 33 | 6.4 | MwHRV | 117 |
| 09/10/2001 | 52.54 | 160.39 | 33 | 5 | mbGS | 133 |
| 10/10/2001 | 52.54 | 160.66 | 33 | 5.4 | MwHRV | 149 |
| 10/10/2001 | 52.61 | 160.48 | 33 | 5.2 | MwHRV | 134 |
| 23/10/2001 | 52.92 | 159.76 | 42 | 5 | mbGS | 76 |
| 15/02/2002 | 52.21 | 159.7 | 33 | 5.1 | MwHRV | 122 |
| 13/04/2002 | 52.99 | 160 | 41 | 5.1 | MwHRV | 91 |
| 08/05/2002 | 52.3 | 160.18 | 44 | 5.5 | MwHRV | 136 |
| 08/05/2002 | 53.81 | 160.77 | 39 | 5.9 | MwHRV | 161 |
| 29/05/2002 | 53.95 | 160.93 | 43 | 5.1 | MwHRV | 178 |
| 20/10/2002 | 52.93 | 160.25 | 46 | 5.1 | MwHRV | 109 |
| 15/03/2003 | 52.25 | 160.39 | 30 | 6.1 | MwGS | 150 |
| 15/03/2003 | 52.42 | 160.13 | 33 | 5.1 | MsGS | 125 |
| 17/03/2003 | 52.36 | 160.12 | 33 | 5.1 | MwHRV | 128 |
| 17/03/2003 | 52.42 | 160.16 | 33 | 5.2 | MwHRV | 126 |
| 17/03/2003 | 52.33 | 160.31 | 33 | 5.5 | mbGS | 141 |
| 18/03/2003 | 52.25 | 160.4 | 33 | 5.4 | MwHRV | 151 |
| 19/03/2003 | 52.13 | 160.78 | 33 | 5.4 | MwHRV | 179 |
| 19/03/2003 | 52.22 | 160.59 | 33 | 5 | mbGS | 163 |
| 19/03/2003 | 52.21 | 160.72 | 33 | 5.6 | MwHRV | 171 |
| 25/03/2003 | 51.99 | 160.57 | 33 | 5 | MwHRV | 179 |
| 08/04/2003 | 53.82 | 160.53 | 33 | 5.1 | MwHRV | 148 |
| 29/05/2003 | 50.88 | 157.3 | 49 | 5.3 | MwHRV | 264 |

Geoacoustic emissions (GAE) in the frequency range from 25 to 1,400 Hz (with transmission coefficient of 0.7 or above relative to the maximum value) have also been recorded in the deep G-1 borehole of the Khlebozavod station under the supervision of V. A. Gavrilov since August 2000. The data analyzed in this paper were obtained by a three-component geophone MAG-3S with magnetoelastic crystal ferromagnetic sensors, designed and developed in the Institute of Physics of the Earth, Russian Academy of Sciences (Belyakov 2000). The vertical channel sensitivity of the geophone is 0.15 V $s^3 m^{-1}$. The sensitivity of horizontal channels is 0.60 V $s^3 m^{-1}$. The output signal of such a sensor is proportional to the third derivative of ground displacement, i.e., jerk, and the gain slope is 60 dB per decade of frequency change. This configuration makes it possible to compensate for the increase in the damping of geoacoustic emissions with frequency in a real geophysical medium. The geophone was set up at the depth of 1,035 m, which is enough to reduce anthropogenic noise levels by more than two orders of magnitude (Gavrilov *et al.* 2008). The geophone body was fixed inside the borehole casing by a spring. The sensor output signals are separated by third-octave band pass filters into four frequency bands with central frequencies 30, 160, 560, and 1,200 Hz, which is followed by the measurement of mean square values of signals from each filter. The resultant signals are then digitally processed: the values of signals are averaged for one-minute intervals and the data are recorded. More detailed description of geoacoustic emission observations and experimental setup for the G-1 borehole is presented elsewhere (Gavrilov *et al.* 2008, Gavrilov *et al.* 2011).

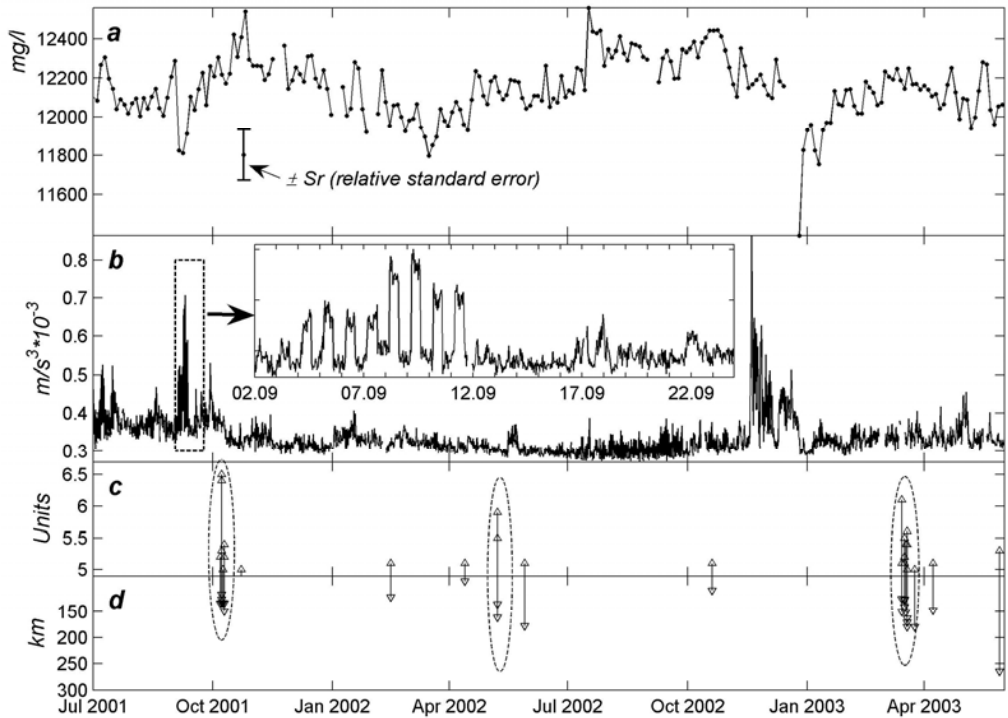

Fig. 2. Experimental data. *a* – variation of water salinity at G-1 borehole (Klebozavod station). Vertical lines with horizontal ticks denote doubled value of relative standard error ($\pm S_r$). *b* – time series of GEA variation at G-1 borehole. Frame *b* also shows an enlarged view of a GAE segment illustrating the degradation of diurnal cycle. *c* and *d* – time sequences of earthquakes expressed in terms of magnitudes and epicenter distances. Dashed lines in panels *c* and *d* denote three groups of strongest seismic events in 2001, 2002, and 2003, respectively.

The objects of this study are GEA time series for central frequency 160 Hz and water salinity times series, both recorded at the G-1 borehole from 30.06.2001 to 1.06.2003 (Fig. 2). The most significant earthquakes (with magnitudes over 5 units) in this time range are listed in Table 1. The water salinity time series is an integral characteristic of the chemical composition of groundwater calculated as a sum of concentrations for basic hydrogeochemical parameters ($HCO_3^-$, $Cl^-$, $Ca^{2+}$, $SO_4^{2-}$, $Na^+$, $K^-$, $Mg^{2+}$). The sum is dominated by the contributions from $Cl^-$-ion (multi-year-averaged value is 6600 mg/l) and $Na^+$-ion (4200 mg/l). This time frame was selected due to the following three reasons. First, there were no large gaps in the GEA measurements during this period. Second, this time frame includes several large earthquakes ($M > 5$) with epicenters located relatively close to the Khlebozavod station (distances in the range from 100 to 200 km). Third, the individual time series, especially for water salinity, do not contain apparent anomalies (precursors) clearly related to earthquakes during this period. These earthquakes were clustered in time into three groups (Fig. 2c,d).

## 4. RESULTS AND DISCUSSION

Hydrogeochemical and geoacoustic observations are two independent methods for monitoring seismotectonic processes that are performed to continuously collect changes in various parameters. The goal of observations is to identify and study effects intrinsically related to the preparation and occurrence of strong earthquakes. The independence of these two types of observations implies that the information on the preparation of a strong earthquake may look differently in the statistical structure of their experimental data. It was previously shown that the most reliable precursor of a strong earthquake on the Kamchatka peninsula is a significant and usually reversible drop or rise of the average concentration of materials dissolved in water (Khatkevich and Ryabinin 2006). The description and analysis of similar hydrogeochemical anomalies as well as irreversible ones, recorded at the G-1 borehole, can be found in the study by Biagi *et al.* (2004). The results of multi-year geoacoustic observations show that in seismically quiescent periods the GAE variation at the G-1 borehole for central frequencies 30 and 160 Hz contains a clear daily component with a maximum at night time ("diurnal cycle"). A suppression (degradation) of the diurnal cycle was suggested as a precursor of strong earthquakes (Gavrilov *et al.* 2008), which is illustrated in Fig. 2b. In this paper, we demonstrate that precursor-related information may be derived from the cross-correlation analysis of hydrogeochemical and geoacoustic data by the FNS method.

　　The combined analysis of hydrogeochemical and geoacoustic data was complicated by the fact that the sampling periods for hydrogeochemical and geoacoustic data were much different: 3 days and 1 minute, respectively. The experimental data could be reduced to the same sampling frequency either by decreasing the frequency for geoacoustic data or increasing it for the water salinity time series. We used a trade-off solution by changing the sampling frequency of both time series to $(6\ h)^{-1}$. This sampling frequency was selected because it is the highest frequency that does not erase information about the diurnal cycle. At the same time, the artificial step-up in the sampling frequency of water salinity data to $(6\ h)^{-1}$ does not distort the information at frequencies $(3\ \text{days})^{-1}$ and below. The complete preprocessing procedure included the removal of single-point spikes and resampling of both time series to uniform sampling frequency $(6\ h)^{-1}$. The thinning of GAE data from sampling frequency $(1\ \text{min})^{-1}$ to $(6\ h)^{-1}$ was performed using the following approach. A low-frequency filter with a cutoff frequency equal to the new Nyquist frequency (12 hr)$^{-1}$ was applied to the GAE data to eliminate the false frequency components. The final GAE time series with the sampling frequency of (6 h)$^{-1}$ was formed by taking every 360$^{th}$ point from the filtered series. The increase of the sampling rate for water salinity time series from (3 days)$^{-1}$ to (6 h)$^{-1}$ was achieved using linear interpolations.

To analyze the mutual dynamics of time series, we used the cross-correlation function (5). The graphical representation of $q(\tau,\theta)$ is a 3D surface, which often has a relatively complex shape. The analysis of cross-correlation evolution in time is performed by calculating $q(\tau,\theta)$ within a sliding window $T$. A time sequence of such 3D surfaces with a shifted sliding window is used to build frames for a video file in the *avi* format.

The analysis of 3D cross-correlation plots for the water salinity and geoacoustic time series allowed us to detect a peculiarity in the structure of correlations prior to the group of earthquakes in October 2001. This peculiarity represents a significant negative correlation between the time series at time lags $\tau$ more than 5 days and time shifts $\theta \approx 0$ days. We observed the same effect prior to the group of earthquakes in March 2003 and saw its less-pronounced instance prior to the group of events in May 2002. Figure 3 illustrates examples of 3D surfaces of cross-correlation $q(\tau,\theta)$ calculated for different time intervals of length $T = 30$ days. 3D plots numbered II, IV, and VI (Fig. 3d) correspond to the intervals preceding the seismic events of 2001, 2002, and 2003. These plots illustrate the peculiarity, i.e., the onset of a major negative correlation with a value of approximately 0.7. The time spans between the right boundaries of intervals II, IV, VI and time moments of strongest earthquakes in 2001, 2002, and 2003 are 27, 31, and 35 days, respectively. It should be pointed out that the data resampling procedure could not significantly distort the observed effect of major negative correlation because its manifestation only at $\tau > 5$ days implies the low contribution of high-frequency components ($f_d = 1$ min$^{-1}$) in GAE signals. Moreover, the functional form of cross-correlation function (5), which is based on differences of the dynamic variables, is determined by the low-frequency components of the signals (Timashev 2006, Timashev and Polyakov, 2007). Let us also note that the absolute value $q_{min}$ of cross-correlation (5) at $\tau > 5$ varied only by 10% when the sampling rate of the GAE signal was gradually reduced in the range from (30 min)$^{-1}$ to (6 h)$^{-1}$ It is clear that the further reduction of sampling frequency to values of ~(10 days)$^{-1}$ causes the major negative cross-correlations to disappear.

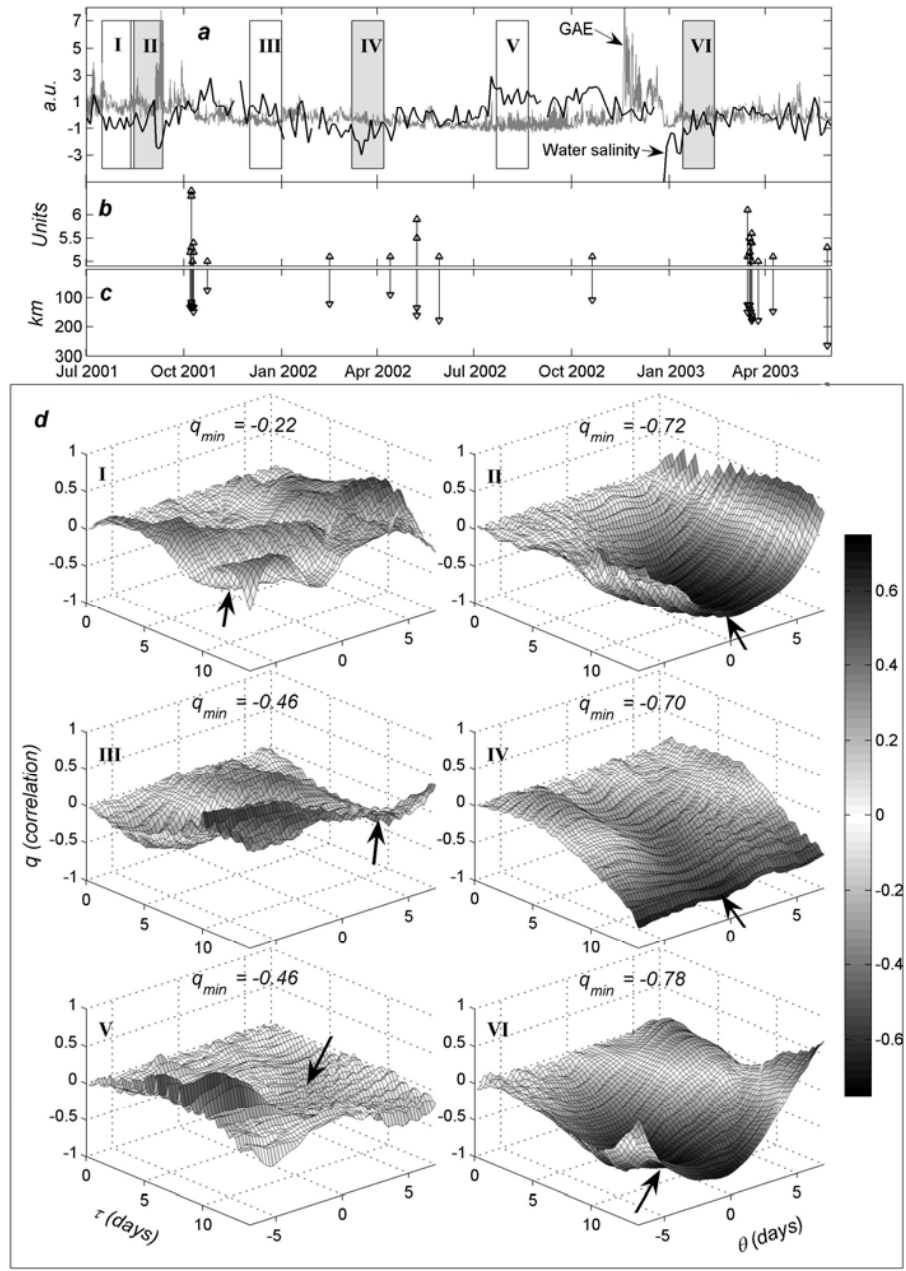

Fig. 3. Results of FNS cross-correlation analysis. *a* – GEA and water salinity time series for G-1 borehole. *b* and *c* – time sequences of earthquakes expressed in terms of magnitudes and epicenter distances. Rectangles in plot *a* denote intervals ($T = 30$ days) used in evaluating the 3D surfaces of cross-correlation $q(\tau, \theta)$ ($\tau_{max} = 14$ days, $\theta_{max} = \pm 7$ days), which are presented in plots *d*. Roman numerals in plots *a* and *d* refer to respective time intervals. Arrows near 3D surfaces denote the positions of correlation minima $q_{min}$, with values of the latter listed above each plot.

The 3D plots of cross-correlation $q(\tau,\theta)$ labeled I, III, and V are shown for comparison and illustrate the cross-correlations for time intervals much further away from the groups of seismic events. All six 3D plots also suggest that the behavior of cross-correlations varies at different intervals. This variability is observed not only in extreme values of cross-correlation $q(\tau,\theta)$, but also in the shape of 3D surfaces. It can be seen that the structure of 3D surfaces on intervals II, IV, and VI, characterized by maximum negative correlations of at least 0.7, has similar features representing a smooth decline of cross-correlation with an increase in time lag $\tau$ and time shift $\theta$ being close to zero. These features in the structure and magnitude of the cross-correlations between water salinity and geoacoustic time series, which were observed in the time frame from 2001 to 2003, allow us to consider them as precursors of seismic events. It should be pointed out that this is a new type of precursor incorporating both the quantitative measure of cross-correlation and its qualitative features described by the structure of 3D surface for cross-correlation $q(\tau,\theta)$.

It should be noted that on certain intervals of length $T=30$ days in the time frame from 2001 to 2003 the structure of 3D surfaces contained high values of negative correlation $q_{min} < -0.7$ between water salinity and geoacoustic emissions, even though those intervals cannot be associated with significant seismic events. Such examples are illustrated in Fig. 4, which shows that the extreme vales of $q_{min}$ took place at values of $\theta$ far from zero. Moreover, the shape of 3D surfaces in Fig. 4 was dramatically different as compared to the monotonous decline of $q(\tau,\theta)$ at $\theta \approx 0$ on intervals II, IV, and VI of Fig. 3. Therefore, the cross-correlation structures plotted in Fig. 4 could not be considered as precursors of strong earthquakes, despite the high values of negative correlation on specific segments of the 3D structures.

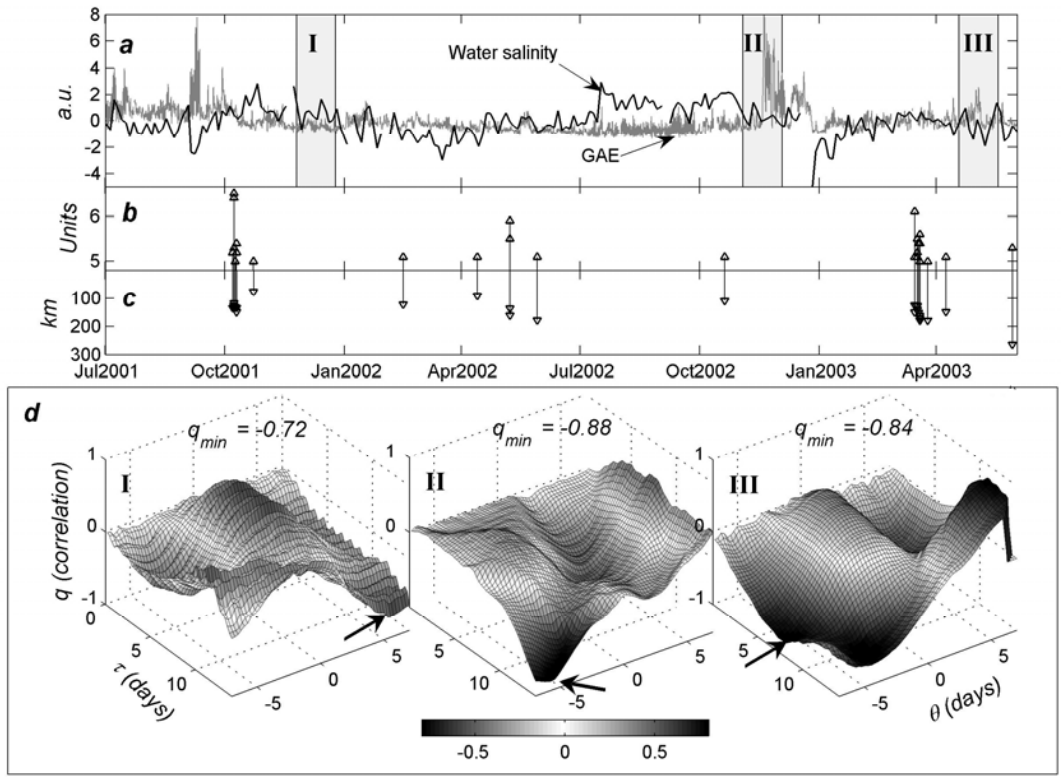

Fig. 4. Results of FNS cross-correlation analysis for non-precursory intervals with high absolute values of correlation minima $q_{min}$. Nomenclature same as in Fig. 3.

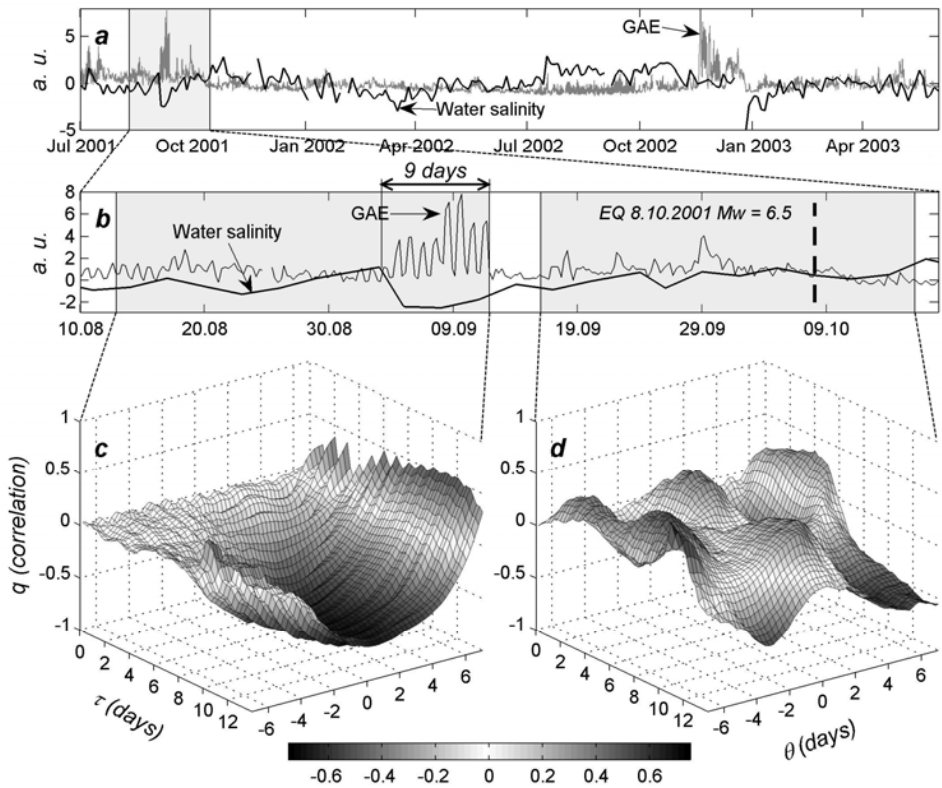

Fig. 5. Plots illustrating the variation of cross-correlation in the time frame around October 2001 (first group of earthquakes). *a* – GEA and water salinity time series for G-1 borehole. *b* – enlarged segment of time series. Time along the abscissa axis in plot *b* is presented in dd.mm format. Vertical dashed line corresponds to the time of earthquake on 8.10.2001. Shaded rectangular areas in plot *b* highlight the intervals of length $T = 30$ days used in evaluating the 3D surfaces of cross-correlation $q(\tau, \theta)$ ($\tau_{max} = 14$ days, $\theta_{max} = \pm 7$ days), which are plotted in *c* and *d*. The horizontal arrow denotes the interval of abrupt increase in GAE (9 days in total) and simultaneous decrease in water salinity.

High values of negative correlation at practically zero values of time shift imply that the changes in water salinity and GAE level happen almost simultaneously, but in opposite directions. To clarify this statement, let us consider in detail the cross-correlation dynamics around the 8.10.2001 earthquake (Fig. 5). It can be seen that the 3D plots of cross-correlation $q(\tau, \theta)$ for two different time intervals vary substantially. In the first case, the cross-correlation is relatively simple and has one clear correlation minimum (Fig. 5c). In the second case, the structure is more complex and contains several local maxima and minima (Fig. 5d). These plots suggest that cross-correlation on the first interval is dominated by simultaneous increase in the GAE level and decline in the water salinity. It can also be seen that the rise in the GAE level can be attributed to the increase in the amplitudes of daily variations (diurnal cycle). On the second interval, which is characterized by a lack of the GAE diurnal cycle, the negative correlations are not significant.

The above results demonstrate that the dynamics of cross-correlations between water salinity and GAE time series contains precursors for all three groups of seismic events in the time frame from 2001 to 2003. The geoacoustic signals generated by a seismically active medium are usually attributed to simultaneous processes of formation/expansion of micro-fractures and their collapse/contraction due to stretching and compression associated with structural reconfiguration of the medium. This also changes the conditions for transport processes in groundwater containing various dissolved inorganic and organic components as well as a variety gases, both in dissolved and microbubble forms. We suggest that the interrelation between fracture expansion (contraction) and increase (decrease) of water permeability in the medium, accompanied by increased (decreased) transport of hydrated ions due to "sieving" and electrokinetic effects (Timashev 1991, Zeman and Zydney 1996, Cicerone *et al.* 2009), may produce the negative correlations shown in Figs. 3d and 5c.

## 5. CONCLUSIONS

This study shows that new precursory information can be extracted from the analysis of the dynamics of cross-correlations between geophysical signals, even when there are no apparent anomalies in the signals themselves. This conclusion is illustrated through the combined analysis of the dynamics for water salinity and geoacoustic emissions in the time frame from 2001 to 2003 at one of the boreholes in a seismically active region of the southeastern part of the Kamchatka peninsula. The individual dynamics of measured times series did not contain characteristic anomalies pointing to possible strong earthquakes, except for the precursor of the October 2001 group of earthquakes in the dynamics of geoacoustic emissions. At the same time, the cross-correlations between water salinity and geoacoustic emissions demonstrated clear precursors prior to all three groups of significant seismic events in this time frame. This result is in agreement with known physical views on the transport of endogenous fluids in rock volumes during the formation of an earthquake source in a seismically active medium, accompanied with fracturing and related geoacoustic phenomena. The new earthquake precursor type suggested in this paper may also be used for the cross-correlation analysis of other known geophysical signals with precursory properties: electromagnetic, hydrological, gasgeochemical, geodetic, and seismic.

Acknowledgements. This study was supported in part by the Russian Foundation for Basic Research, grant nos. 09-05-98543-r_vostok_a, 11-02-00540 a, 10-02-01346 a, 12-05-00670 a.

References

Bella, F., P.F. Biagi, M. Caputo, E. Cozzi, G.D. Monica, A. Ermini, E.I. Gordeev, Y.M. Khatkevich, G. Martinelli, W. Plastino, R. Scandone, V. Sgrigna, and D. Zilpimiani (1998), Hydrogeochemical anomalies in Kamchatka (Russia), *Phys. Chem. Earth* **23**, 921–925, DOI: 10.1016/S0079-1946(98)00120-7.

Belyakov, A.S. (2000), Magnetoelastic acoustic geophones for geophysical research and earthquake prediction (in Russian), *Seismicheskie Pribory* **33**, 27–45.

Biagi, P.F., A. Ermini, E. Cozzi, Y.M. Khatkevich, and E.I. Gordeev (2000), Hydrogeochemical precursors in Kamchatka (Russia) related to the strongest earthquakes in 1988–1997, *Nat. Hazards* **21**, 263–276, DOI: 10.1023/A:1008178104003.

Biagi, P.F., L. Castellana, A. Minafra, G. Maggipinto, T. Maggipinto, A. Ermini, O. Molchanov, Y.M. Khatkevich, and E.I. Gordeev (2006), Groundwater chemical anomalies connected with the Kamchatka earthquake (M=7.1) on March 1992, *Nat. Hazards Earth Syst. Sci.* **6**, 853–859, DOI: 10.5194/nhess-6-853-2006.

Biagi P.F., L. Castellana, R. Piccolo, A. Minafra, G. Maggipinto, A. Ermini, V. Capozzi, G. Perna, Y.M. Khatkevich, and E.I. Gordeev (2004), Disturbances in groundwater chemical parameters related to seismic and volcanic activity in Kamchatka (Russia), *Nat. Hazards Earth Syst. Sci*. **4**, 535–539, DOI: 10.5194/nhess-4-535-2004.

Cicerone, R.D., J.E. Ebel, and J. Britton (2009), A systematic compilation of earthquake precursors, *Tectonophysics* **476**, 371–396, DOI: 10.1016/j.tecto.2009.06.008.

Fedotov, S.A., A.A. Gusev, L.S. Shumilina, and V.G. Chernyshova (1985), The seismofocal zone of Kamchatka (in Russian), *Vulkanologiya i Seismologiya* (4), 91–107.

Gavrilov, V., L. Bogomolov, Y. Morozova, and A. Storcheus (2008), Variations in GAE in a deep borehole and its correlation with seismicity, *Ann. Geophys.* **51**, 737–753, DOI: 10.4401/ag-3013.

Gavrilov V., L. Bogomolov, and A. Zakupin (2011), Comparison of the geoacoustic measurements in boreholes with the data of laboratory and in-situ experiments on electromagnetic excitation of rocks, *Izv. Phys. Solid Earth* **47**, 1009–1020.

Geller, R.J. (1997), Earthquake prediction: a critical review, *Geophys. J. Int.* **131**, 425–450, DOI: 10.1111/j.1365-246X.1997.tb06588.x.

Geller, R.J., D.D. Jackson, Y.Y. Kagan, and F. Mulargia (1997), Earthquakes cannot be predicted, *Science* **275**, 1616–1617, DOI: 10.1126/science.275.5306.1616.

Hartmann, J. and J.K. Levy (2005), Hydrogeological and gasgeochemical earthquake precursors - A review for application, *Nat. Hazards* **34**, 279–304, DOI: 10.1007/s11069-004-2072-2.

ICEFCP (2009), Operational Earthquake Forecasting: State of Knowledge and Guidelines for Utilization, http://www.iaspei.org/downloads/Ex_Sum_v5_THJ9_A4format.pdf, access: 16 September 2011.

Khatkevich, Y. and G. Ryabinin (2006), Geochemical an ground-water studies in Kamchatka in the search for earthquakes precursors (in Russian), *Vulkanologiya i Seysmologiya* (4), 34–42.

Kopylova, G., V. Sugrobov, and Y. Khatkevich (1994), Variations in the regime of springs and hydrogeological boreholes in the Petropavlovsk polygon (Kamchatka) related to earthquakes (in Russian), *Vulkanologiya i Seysmologiya* (2), 53–70.

Lyubushin, A.A., G.N. Kopylova, and Yu.M. Khatkevich (1997), Analysis of the spectral matrices of hydrogeological observations at the Petropavlovsk geodynamic research site, Kamchatka, and their comparison with the seismic regime, *Izv. Phys. Solid Earth* **33**, 497-507.

Lyubushin, A.A. (1998), An aggregated signal of low-frequency geophysical monitoring systems, *Izv. Phys. Solid Earth* **34**, 238-243.

Lyubushin, A.A. (2000), Wavelet-aggregated signal and synchronous peaked fluctuations in problems of geophysical monitoring and earthquake prediction, *Izv. Phys. Solid Earth* **36**, 204-213.

Lyubushin, A.A (2007), *Analiz Dannykh Sistem Geofizicheskogo i Ekologicheskogo Monitoringa* (Data Analysis for Systems of Geophysical and Environmental Monitoring), Nauka, Moscow.

Ryabinin, G., Yu. S. Polyakov, V. A. Gavrilov, and S. F. Timashev (2011), Identification of earthquake precursors in the hydrogeochemical and geoacoustic data for the Kamchatka peninsula by flicker-noise spectroscopy, *Nat. Hazards Earth Syst. Sci.* **11**, 541-548, DOI: 10.5194/nhess-11-541-2011.

Telesca, L., M. Lovallo, A. Ramirez-Rojas, and F. Angulo-Brown (2009a), A nonlinear strategy to reveal seismic precursory signatures in earthquake-related self-potential signals, *Physica A* **388**, 2036–2040, DOI: 10.1016/j.physa.2009.01.035.

Telesca, L., M. Lovallo, A. Ramirez-Rojas, and F. Angulo-Brown (2009b), Scaling instability in self-potential earthquake-related signals, *Physica A* **388**, 1181–1186, DOI: 10.1016/j.physa.2008.12.029.

Timashev, S. F. (1991), Physical Chemistry of Membrane Processes, Ellis Horwood, Chichester.

Timashev, S. F. (2006), Flicker noise spectroscopy and its application: Information hidden in chaotic signals, *Russ. J. Electrochem.* **42**, 424-466, DOI: 10.1134/S102319350605003X.

Timashev, S. F. (2007), *Fliker-Shumovaya Spektroskopiya: Informatsiya v khaoticheskikh signalakh* (Flicker-Noise Spectroscopy: Information in Chaotic Signals), Fizmatlit, Moscow.

Timashev, S. F. and Y.S. Polyakov (2007), Review of flicker noise spectroscopy in electrochemistry, *Fluct. Noise Lett.* **7**, R15-R47, DOI: 10.1142/S0219477507003829.

Timashev, S.F., and Y. S. Polyakov (2008), Analysis of discrete signals with stochastic components using flicker noise spectroscopy, *Int. J. Bifurcation Chaos* **18**, 2793-2797, DOI: 10.1142/S0218127408022020.

Timashev, S. F., Y.S. Polyakov, P.I. Misurkin, and S.G. Lakeev (2010), Anomalous diffusion as a stochastic component in the dynamics of complex processes, *Phys. Rev. E* **81**, 041128, DOI: 10.1103/PhysRevE.81.041128.

Wyss, M. and D.C. Booth (1997), The IASPEI procedure for the evaluation of earthquake precursors, *Geophys. J. Int.* **131**, 423–424, DOI: 10.1111/j.1365-246X.1997.tb06587.x.

Wyss, M., R.L. Aceves, S.K. Park, R.J. Geller, D.D. Jackson, Y.Y. Kagan, and F. Mulargia (1997), Cannot earthquakes be predicted?, *Science* **278**, 487–490, DOI: 10.1126/science.278.5337.487.

Uyeda, S., T. Nagao, and M. Kamogawa (2009), Short-term earthquake prediction: Current status of seismo-electromagnetics, *Tectonophysics* **470**, 205–213, DOI: 10.1016/j.tecto.2008.07.019.

Zeman, L. J. and A.L. Zydney (1996), *Microfiltration and Ultrafiltration: Principles and Applications*, Marcel Dekker, New York.